\documentstyle[prb, aps, amssymb
 ]{revtex}
\begin{document}
\draft
\wideabs{
\title{Spin polarization contrast observed in GaAs
by force-detected nuclear magnetic
resonance}
\author{Kent R. Thurber}
\address{U.S. Army Research Laboratory, AMSRL-SE-EM, Adelphi, Maryland
20783 and
Center for Superconductivity Research, Department of Physics, University of
Maryland,
College Park, Maryland 20742}
\author{Lee E. Harrell}
\address{Department of Physics, U.S. Military Academy, West Point, New York 10996}
\author{Ra\'{u}l Fainchtein}
\address{Johns Hopkins University Applied Physics Laboratory, 11100 Johns
Hopkins Road,
Laurel, Maryland 20723}
\author{Doran D. Smith}
\address{U.S. Army Research Laboratory, AMSRL-SE-EM, Adelphi, Maryland 20783}
\date{\today}
\maketitle
\begin{abstract}
We applied the recently developed technique of force-detected nuclear magnetic
resonance (NMR) to observe $^{71}$Ga, $^{69}$Ga, and $^{75}$As in
GaAs.  The nuclear spin-lattice relaxation time is 21$\pm$5 min for $^{69}$Ga at
$\sim 5$ K
and 4.6 Tesla.  We have exploited this long relaxation time
to first create and then observe spatially varying nuclear spin polarization
within the sample, demonstrating a new form of contrast for magnetic
resonance force microscopy (MRFM).  Such nuclear spin contrast could
be used to indirectly image electron spin polarization in GaAs-based spintronic
devices.
\end{abstract}
}

Due to its long spin coherence lifetime and scattering
length,\cite{Kikkawa,Hagele} GaAs has received attention as a
potential material for spintronic devices that exploit the spin
degree of freedom of the electron.\cite{Ball}  For example, B.
Jonker {\it et al.}\cite{Jonker} have demonstrated a scheme for
the injection of spin-polarized carriers into GaAs/AlGaAs
heterostructures via an epitaxial layer of magnetic ZnMnSe.  A
technique that can image the electron spin polarization at any
arbitrary point in the device would be very useful for determining
whether spin flips are occurring at an interface or in the bulk
material.  Short device transit times make direct imaging of the
electron polarization difficult.  However, dynamic nuclear
polarization\cite{Berkovits,Dobers} provides an indirect route to
the time-averaged electron polarization.  In dynamic nuclear
polarization, carriers become trapped on defects and transfer
their polarization to nearby nuclei via the hyperfine interaction.
The induced nonequilibrium polarization of the nuclear lattice
persists for minutes, so it could be measured by a magnetic
resonance imaging technique with adequate sensitivity and
resolution, such as MRFM,\cite{Sidles} a scanned probe microscopy
based on force-detected magnetic resonance.

In this letter, we demonstrate the first steps towards indirect
electron polarization imaging using force-detected NMR.  We create
and then observe nuclear spin polarization contrast in GaAs.
Figure 1 shows force-detected NMR observations of $^{71}$Ga,
$^{69}$Ga, and $^{75}$As, increasing the number of observed
isotopes from four ($^{1}$H, $^{19}$F, $^{59}$Co,
$^{23}$Na)\cite{Rugar,Wago,Pelekhov,Verhagen} to seven. Further,
this demonstrates the observation of multiple isotopes within a
single sample using force-detected NMR techniques.  In addition,
we used off resonance cantilever excitation, which allows the
measurement of nuclear isotopes with small gyromagnetic ratio,
$\gamma$, without requiring large radio frequency (RF) magnetic
fields.

In force-detected NMR, the sample is mounted on a micro-cantilever in an applied
magnetic field with a magnetic field gradient arising from a nearby magnetic
particle.
The field gradient exerts a force on the magnetized sample.  The magnetization
of a
particular nuclear isotope is isolated by using adiabatic rapid passage
(ARP)\cite{Slichter} to
cyclically invert its magnetic moment at a frequency $f_{ARP}$.  The signal
appears as a driven
cantilever oscillation.  The magnetic field gradient restricts the NMR resonance
to a sensitive slice of the sample, so magnetic resonance imaging is
possible.\cite{Zuger}

The sample and cantilever form a mechanical resonator with resonance frequency
$f_{c}$ and
quality factor $Q$.  The amplitude $A$ of the cantilever response depends on
$f_{ARP}$ and the
amplitude $F$ of the driving force.  For $f_{ARP} \ll f_{c}$, the cantilever
response is essentially the
zero frequency response, $A = F/k$, where $k$ is the spring constant of the
cantilever.  When
driven on resonance, the cantilever response is enhanced by a factor of $Q$.
Well-understood restrictions limit the frequency at which spins can be flipped
by ARP.  In
particular,\cite{Wago}
\begin{equation}
\Omega \leq \frac{\left( \gamma B_{1} \right)^{2}}{4f_{ARP}}
\label{ARPeq}
\end{equation}
where $\gamma$ is the gyromagnetic ratio, and $\Omega$ is the
amplitude of the frequency modulation of the radio frequency (RF)
magnetic field $B_{1}$.  On resonance excitation of the cantilever
($f_{ARP} = f_{c}$) can require a large RF field $B_{1}$ if
$\gamma$ is small.  Our experiment demonstrates the utility of off
resonance cantilever excitation ($f_{ARP} < f_{c}$).  By
eliminating the need to satisfy the adiabatic rapid passage
condition at the cantilever resonance frequency, we reduced our
required RF magnetic field by almost a factor of 8 (and thus
reduced our RF power needs by a factor of 60!)

Our sample is a 3 $\mu$m epilayer of GaAs, doped at 0.6x10$^{18}$
cm$^{-3}$ Si and 2.0x10$^{18}$ cm$^{-3}$ Be, grown on a AlAs
lift-off layer and released using HF.  A hand-cleaved irregular
rectangle of this material ($\approx$210 $\mu$m x 150 $\mu$m) is
attached to the cantilever (k $\sim$ 0.05 N/m)\cite{cantilevers}
using silver-filled epoxy. The cantilever is coated with 0.2
$\mu$m of Au on each side to enhance thermal conductivity. The
loaded cantilever has $f_{c} = 1.034$ kHz and $Q = 154$ at 5 K in
the presence of He exchange gas.  The cryostat and detection
electronics are described in Ref. 15.  The magnetic gradient
source is a 250 $\mu$m diameter Fe cylinder positioned $\sim 35$
$\mu$m from the sample.  Because the sample is thin, the sensitive
slices are circular arcs that increase in radius with applied
magnetic field at fixed RF frequency as shown in Fig. 2(a).

A tuned and matched 1 1/2 turn, 710 $\mu$m diameter coil generates the 4 gauss
transverse RF
magnetic field.  The RF waveform is generated by a Hewlett-Packard model HP8642B
source driven by a $f_{ARP} = 16.6$ Hz triangle wave frequency modulation,
with subsequent
amplification.  The triangle wave is offset such that only positive frequency
deviations up
to 88 kHz (i.e. $\Omega = 44$ kHz) are obtained.  The motion of the cantilever
is detected by a
temperature-tuned fiber optic interferometer\cite{Bruland} with the fiber rigidly
epoxied to the
cantilever support.  The signal is demodulated with a lock-in amplifier referenced
to the
frequency modulation signal.

Figure 1 displays a force-detected NMR spectrum of our sample observed by sweeping
the applied magnetic field at a fixed RF center frequency of 51.56 MHz.  All three
isotopes are clearly resolved with cantilever displacements of $\sim 1.3$
picometers.  The large
width of the NMR peaks represents the spatial extent of the sample.  As
illustrated in Fig.
2(a), each point in the NMR line of each isotope represents the signal from a
ring of the
sample.  On the axis of the cylindrical iron magnet, the magnetic field is
strongest, which
means the applied external magnetic field required for magnetic resonance is
less.  Thus,
the points at lower applied magnetic field represent the sample near the center
of the
magnet.

Using the observed maximum offset of the signal (0.8 Tesla) from the nominal
resonance
fields and the saturation magnetization for iron (2.18 Tesla),\cite{Bozorth}
we can calculate the
magnetic field gradient and the geometry of the sensitive slice.  As an
example, for the
region shown in Fig. 2(a), the volume of a typical sensitive slice is
870 $\mu$m$^{3}$ with a
magnetic field gradient of 8600 T/m.  Using thermal polarization of $^{69}$Ga
at 5 K gives an
estimated 3.4 pm cantilever excitation over the 10 second lock-in
integration period.  Analysis of $^{75}$As and $^{71}$Ga gives similar results.
Our experimental signal is $\sim 2.5$ times smaller than
this estimate.  Recent tests have suggested that RF noise may still be a
problem.

In our 0.1 Hz bandwidth, we observe 0.2 pm rms cantilever noise.  Assuming a
frequency
independent force spectral power density,\cite{Sidles} the root mean
square thermal excitation of the cantilever in
a given narrow bandwidth $\Delta f$ well below resonance,
\begin{equation}
x_{{\rm rms,thermal}} = \left( \frac{2k_{B} T \Delta f}{\pi k
f_{c} Q} \right)^{1/2} \label{thermaleq}
\end{equation}
is smaller than the on resonance thermal noise by a factor of $Q$.
Interestingly, the
observed signal is also a factor of $Q$ smaller in the low frequency
regime relative to an on
resonance excitation, so the thermal signal-to-noise ratio is independent
of $f_{ARP}$.\cite{Brulandthesis}  Using
equation (2) we estimate $x_{{\rm rms,thermal}} = 0.02$ pm.  Currently, our
noise is dominated by the interferometer electronics.

By averaging lock-in traces, we were able to see the time
dependent decay, $\exp(-t/\tau_{m})$,\cite{Leskowitz} of the
nuclear magnetization when cyclically flipped by our ARP sequence.
The observed $\tau_{m}$ values are 7 s for $^{75}$As (applied
fields 6.28 to 6.44 Tesla), 20 s for $^{69}$Ga (4.28 to 4.44
Tesla), and 30 s for $^{71}$Ga (3.26 to 3.42 Tesla) for the data
displayed in Fig. 1.  In our experiment, the driven relaxation
time of the nuclear magnetization depends on the adiabatic
condition for the spin flips (Eq. 1).  Larger $\gamma$ and $B_{1}$
and smaller $f_{ARP}$ increase the relaxation time, $\tau_{m}$.

For future work, we are interested in the nuclear spin-lattice
relaxation time, $T_{1}$, of our sample.  To determine $T_{1}$ for
$^{69}$Ga, we destroyed the magnetization by applying our ARP
sequence for 5 $\tau_{m}$, and then allowed it to recover at an
applied field of 4.0 T (total field 4.5 to 4.7 T) for a range of
delay times before observing the NMR signal.  Fitting the data to
a single exponential gives $T_{1} = 21\pm5$ min, in agreement with
previously published results.\cite{Dobers}

The relatively long $T_{1}$ of GaAs at low temperatures allows us
to create and then measure spin polarization contrast in the
sample before the spins relax back to thermal equilibrium.  Figure
2(b) shows an example of spin polarization contrast created by
using a single ARP sweep to invert the nuclear polarization in all
but a narrow slice of the sample.  In the applied field range 4.39
to 4.45 Tesla, the $^{69}$Ga spin polarization is still aligned
with the magnetic field.  Outside of this field range, the force
signal is negative relative to the thermal signal, indicating that
the magnetization has been flipped to point opposite to the
applied magnetic field.  Figure 2(a) illustrates the spatial
dependence of the corresponding spin polarization contrast in the
sample.  This data demonstrates conclusively that cycling spins by
ARP in one volume of the sample does not disturb the polarization
in nearby out of resonance sample volumes, an important
consideration for efficient imaging of nonequilibrium nuclear
polarization.

In conclusion, we applied force-detected NMR techniques to GaAs,
demonstrating force-detection of spin polarization contrast.  A
spin-lattice relaxation time of 21$\pm$5 min was
observed for $^{69}$Ga.  This work significantly increases
the total number of isotopes
observed and reported in the force-detected NMR literature.  We used
off cantilever
resonance force-detected NMR, which allows the measurement of
nuclei with small $\gamma$ at
reasonable RF fields.  In fact, $^{75}$As has the smallest $\gamma$
measured by force-detected NMR to date.

This work was supported in part by the U.S. Army Research Laboratory Microelectronics
Research Collaboration Program, Grant No. DAAL01-95-2-3530, the DARPA Defense
Science Office, and the U.S. Military Academy/Army Research Laboratory Faculty
Research Program.   The sample was grown and processed by Peter Newman and Monica
Taysing-Lara.  We would like to thank John A. Marohn, James G. Kempf, and
Daniel Weitekamp for many helpful discussions and suggestions.


%
%

\begin{figure}
\caption{Force-detected NMR signal of GaAs as a function of
applied magnetic field at constant RF center frequency 51.56 MHz.
Spins were flipped at $f_{ARP} = 16.6$ Hz with ARP frequency
modulation of $\pm$44 kHz.  Lock-in detection was used with a 1
s time constant, and 10 s of data were integrated.
$^{71}$Ga, $^{69}$Ga, and $^{75}$As are well resolved. Peak widths
are largely determined by the spatial extent of the sample and the
magnetic field gradient.  Offset of the signal from the nominal
Larmor frequency is due to the extra magnetic field of the
gradient source.  A linear background has been subtracted from the
data for clarity. } \label{fullscan}
\end{figure}

\begin{figure}
\caption{(a) Illustration of the induced nuclear spin polarization
contrast (sample, cantilever and iron cylinder (magnet) drawn to
scale - sample and cantilever thickness exaggerated).  The
interferometer fiber and RF coil are omitted for clarity. The
circular arc on the sample indicates the 11 $\mu$m wide region
where the $^{69}$Ga spin polarization is aligned with the magnetic
field, while the $^{69}$Ga spins in the rest of the sample are
aligned opposite the applied magnetic field. (b) Force-detected
$^{69}$Ga NMR signal as a function of applied magnetic field.  The
thermal polarized data ($\square$) is shown expanded from Fig. 1.
The second set of data ($\bullet$) was measured after the
$^{69}$Ga spin polarization was patterned by an ARP sweep.  A
single ARP sweep inverted the nuclear polarization for most of the
sample except for the applied field region 4.39 to 4.45 Tesla when
the RF field was turned off (illustrated in part a).  In order to
enhance the contrast, the nuclear spins were thermalized at 8
Tesla prior to patterning, the lock-in was used with a 0.3 s time
constant, and 5 s of data was integrated.} \label{contrast}
\end{figure}

%
%

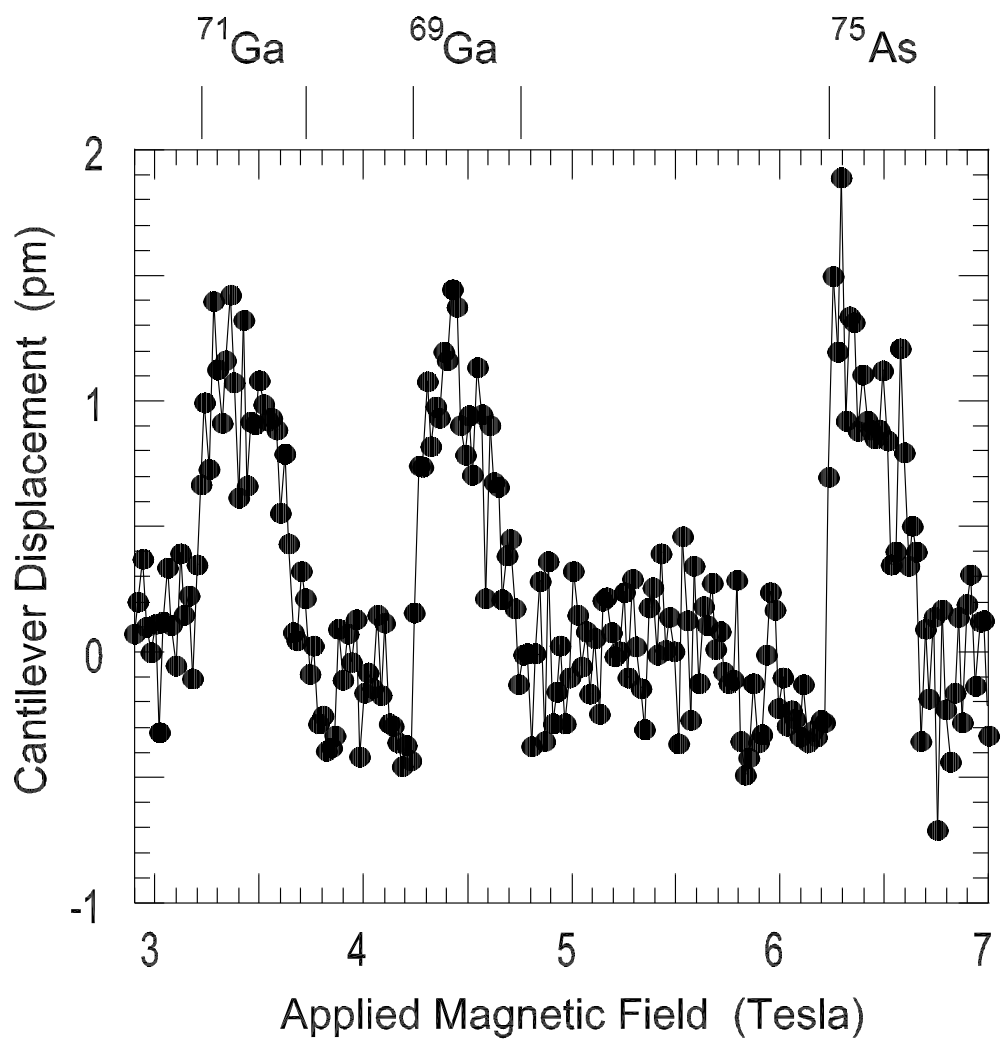
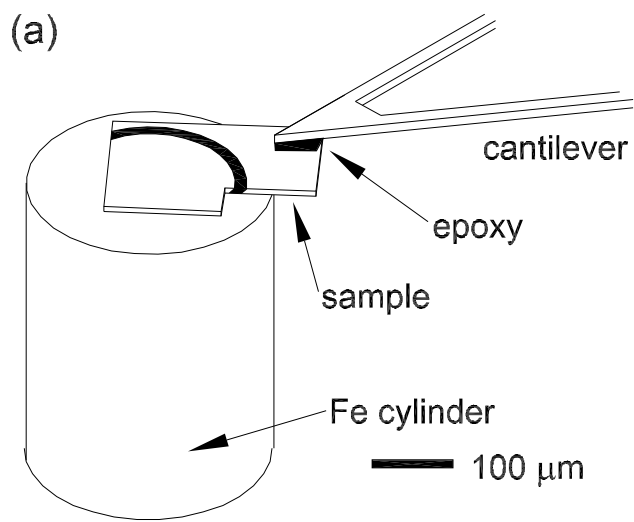
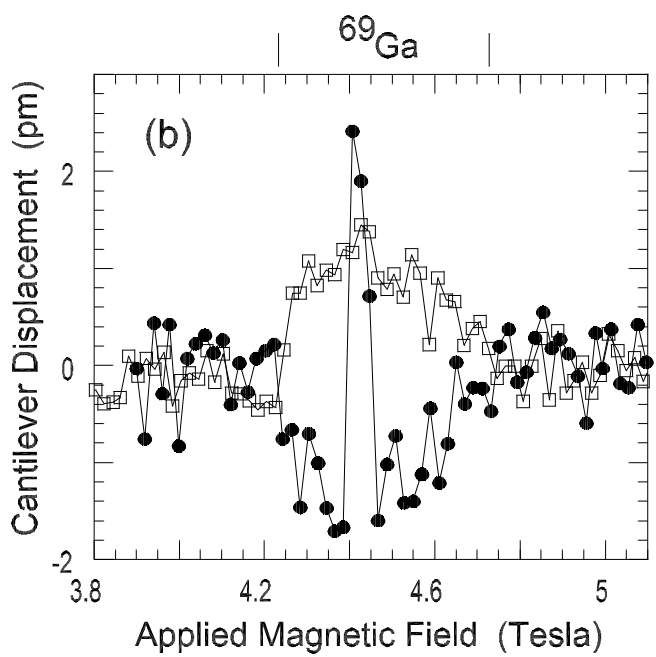
\end{document}